# Nitrogen-Vacancy Engineering for Controlled Phase Transitions in CrN(111) Epitaxial Films

XiaoXu Zhang, Yang Li, Yu Shang, MingYue Zhao, GuoKe Li[*], Li Ma, DeWei Zhao, CongMian Zhen, DengLu Hou

*Hebei Key Laboratory of Photophysics Research and Application, College of Physics, Hebei Normal University, Shijiazhuang, 050024, China.*

**Abstract:** The phase transition in CrN epitaxial films is substantially suppressed by epitaxial constraint. Here, we propose that nitrogen (N) vacancies can be taken as a knob to regulate the phase transition of CrN(111) epitaxial films. To validate this concept, a series of CrN(111) films with controlled N concentrations (approximately from 0.0 to 5.0 at.%) were epitaxially grown on $Al_2O_3$(0001) substrates. Experimental characterization reveals that higher N vacancy concentrations significantly facilitate the out-of-plane contraction of the films at 273 K (0.8%), reaching up to 60% of the contraction magnitude of CrN powders (1.2%) without compromising the stability and reproducibility of the phase transition. Reducing N vacancy concentrations diminishes the lattice contraction, lowers the phase transition temperature to 193 K, and triggers a metallic to insulator transition in electrical behavior. First-principles calculations corroborate these findings, showing that N vacancies decrease the internal tensile stress within triangular Cr atomic layers, which enhances the out-of-plane contraction, elevates phase transition temperatures, and promotes bandgap closure. These results establish N vacancies as a critical factor governing phase transition dynamics in CrN systems and provide a practical strategy for successively engineering thermally responsive phase transitions in CrN films, advancing their potential for functional device applications.

---

[*] Corresponding author. E-mail address: liguoke@126.com.

# 1. Introduction

Bulk CrN powders transition from a paramagnetic metallic rock salt structure to a [110]$_2$-type antiferromagnetic insulating orthorhombic (Pnma) structure below a Néel temperature ($T_N$) of 286 K [1-3]. This triple-phase transition, arising from strong electron-spin-lattice coupling, renders CrN highly sensitive to external stimuli like strain and doping [2-7]. These properties, along with a narrow band gap [8-14], position CrN as a promising material for thermoelectric [4, 5, 15], information storage [16], and antiferromagnetic spintronics [17]. Transition metal nitrides, including CrN, offer ultra-hardness, high melting points, and excellent corrosion and oxidation resistance [18, 19], ensuring stability and environmental adaptability. CrN films are particularly valuable due to their reduced dimensionality and tunable physical properties [15, 16]. However, structural phase transitions and related electronic transport behaviors in thin films are significantly suppressed due to dimensional reduction and epitaxial constraints [20], highlighting the need for strategies to optimize these transitions for novel device applications.

In contrast to the pronounced resistivity jump in CrN bulk powders, CrN films usually show minor discontinuities or continuous resistivity across their transition temperature (TN), indicating suppressed phase transitions [13, 21, 22]. These films exhibit resistivity spanning five orders of magnitude at room temperature, encompassing metal-to-metal, insulator-to-metal, and insulator-to-insulator transitions [13, 21, 22], with transition temperatures ranging from 290 K to 80 K [5, 6, 8, 9, 15-17, 21-39]. This variability suggests potential for external control over CrNs phase transitions, enhancing its applicability, but also reveals gaps in understanding the underlying mechanisms. Factors such as N vacancies[40-44], point defects[5, 40], epitaxial strain [21, 33, 42, 45], crystallinity, film thickness[16], orientations[17, 37], grain boundaries [21, 33], surface states[36, 46], and secondary phases[38], are proposed to contribute to these variations. Given that CrN tends to be off stoichiometry and its resistivity is highly dependent on nitrogen vacancy concentrations[41], we compiled data illustrating the relationship between structural phase transition temperature and resistivity in CrN films (Fig. S1) [5, 6, 8, 9, 15-17, 21-39]. It reveals that structural phase transitions occur exclusively at lower resistivity levels, underscoring the decisive role of nitrogen vacancies. Notable phase transition behaviors occasionally observed in CrN films with extremely low resistivity suggest potential strategies for regulating structural phase transitions by controlling nitrogen vacancies [16, 26]. Therefore, this study aims to explore the influence of N vacancies on the phase transition of CrN (111) epitaxial films and to experimentally manipulate these vacancies to control the transition.

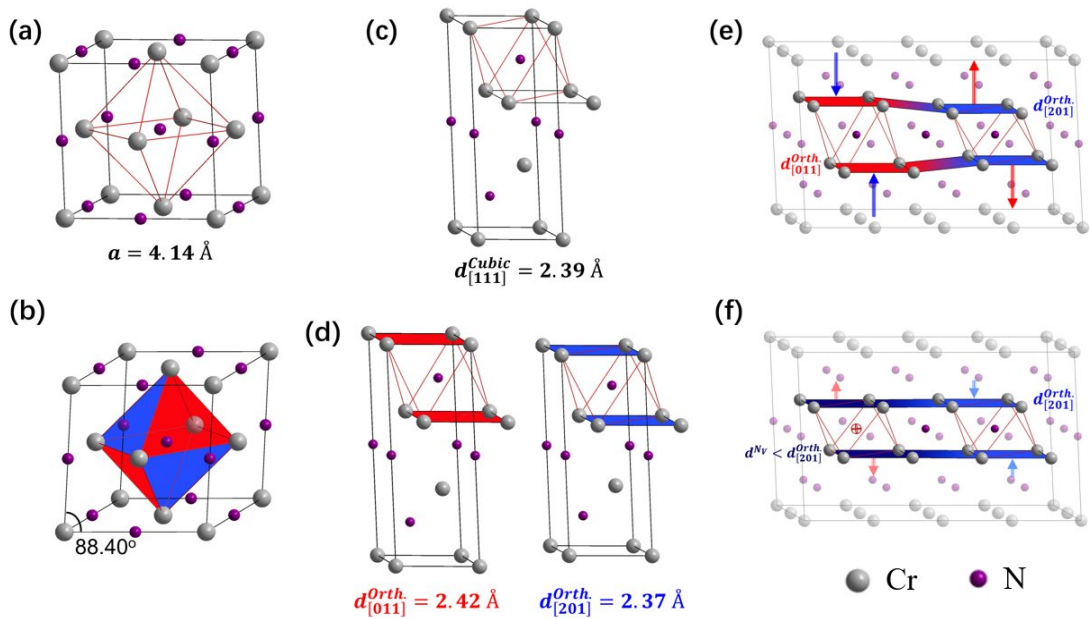

**Fig. 1** Schematic of N vacancies effect on CrN(111) epitaxial films structural phase transition. (a) Cubic CrN unit cell with rock salt structure above $T_N$. (b) a distorted cubic CrN unit cell below $T_N$, showing orthorhombic CrN(011) and (201) planes in red and blue. (c) Along [111], cubic CrN has alternating triangular Cr/N layers, 2.39 Å apart. (d) Orthorhombic CrN projections along [011] and [201], with triangular Cr layer spacings of 2.42 Å and 2.37 Å. (e) Below $T_N$, two domain types with distinct spacings emerge, causing boundary stress that hinders phase transition. (f) N vacancies induce film contraction along the normal direction, reducing internal stress and promoting phase transition; arrows depict stress magnitude and direction.

To begin with, we examine how epitaxial constraints influence the structural phase transition in an ideal CrN(111) epitaxial film. In Fig. 1(b), cubic CrN consists of alternating triangular Cr and N atomic layers along the [111] direction with a 2.39 Å interplanar spacing. Below $T_N$, these planes equally transform into orthorhombic CrN{011} and {201} planes (Fig. 1(c)), resulting in slightly distorted triangular Cr layer spacings of 2.42 Å and 2.37 Å, respectively (Fig. 1(d)) [1, 3]. For CrN(111) films, substrate constraints restrict the movement of Cr atomic layers parallel to the surface. Perpendicular to the film surface, two distinct types of domains corresponding to orthorhombic CrN{011} and {201} planes would emerge below $T_N$, as shown in Fig. 1(d). Owing to their differing interplanar spacings, these domains induce internal stress at their boundaries (Fig. 1(e)), significantly hindering the structural distortion. In this context, alleviating internal stress through nitrogen vacancies, which generate chemical stress that contracts with Cr atoms [44], can facilitate the structural phase transition. As illustrated in Fig. 1(f), this chemical stress would reduce domain numbers with increased spacing, thereby promoting the structural phase transition by reducing internal tensile stress. Consequently, the phase transition would be highly sensitive to N vacancy concentration, making it possible to successively control the structural phase transition in CrN(111) epitaxial films by manipulating N vacancy concentrations.

To verify the role of N vacancies, a series of CrN(111) epitaxial films were grown on $Al_2O_3$(0001) substrates. The choice of $Al_2O_3$(0001) substrates emphasizes the impact of N vacancies, as achieving the structural phase transition of CrN on this substrate is particularly challenging [28, 37]. The $N_2$ gas flow rate was the sole variable parameter, incrementally adjusted to finely tune the N vacancy concentration within the films. Subsequently, we conducted a systematic characterization of the structural phase transitions and electrical transport properties of these films to elucidate the role of N vacancies.

## 2. Experimental methods

Chromium nitride films were deposited on $Al_2O_3(0001)$ substrates using DC magnetron sputtering from a high-purity Cr target (99.95%). The base pressure was below $1.0\times10^{-5}$ Pa. A constant total gas flow rate of 50.0 sccm ($Ar + N_2$) was maintained at a sputtering pressure of 1.0 Pa. N vacancy concentration was adjusted by varying the $N_2$ gas flow rate ($F_{N2}$) from 4.0 to 16.0 sccm. Films were labeled according to their $F_{N_2}$ values; for example, a film prepared at 6.0 sccm was denoted as $F_{6.0}$. The sputtering power and substrate temperature were fixed at 60.0 W and 700°C, respectively. Substrate rotation at 3.0 rpm ensured uniform film thickness. After 30 minutes of sputtering, films reached a thickness of 220 nm, confirmed by cross-sectional transmission electron microscopy. All measurements were performed on this consistent set of samples without exception.

The crystallographic characteristics of the films were analyzed using X-ray diffraction (XRD) with a Philips XPert Pro system, complemented by a low-temperature attachment. Off-axis azimuthal φ-scans were employed to investigate the in-plane epitaxial relationships, while ω-rocking curve scans were utilized to assess the crystalline quality. Variable temperature XRD measurements were conducted to explore variations in out-of-plane spacing with temperature, during which the effect of Al2O3(0001) substrates was neglected for all samples. Considering the limited number of temperature data points available, phase transition temperatures were determined by averaging the two adjacent temperature points exhibiting the highest rate of change in out-of-plane spacing.

X-ray photoelectron spectroscopy (XPS) measurements were conducted using a Thermo ESCALAB 250Xi with monochromatized Al-K$_\alpha$ radiation (1486.6 eV). The base pressure during analysis was approximately $5 \times 10^{-7}$ Pa. No sputter etching was performed before XPS measurements. Three sets of N 1$s$ and Cr 2$p$ core level spectra were measured in different areas of $0.5 \times 0.5$ mm$^2$ for each film. All XPS peaks were calibrated for adventitious carbon with a C1$s$ peak binding energy of 284.80 eV. The atomic ratio of the films was calculated as the average of the atomic ratios of N to Cr obtained by deconvolution of these three sets of N1$s$ and Cr 2$p$ spectra.

The microstructure of the film deposited at $F_{N_2} = 6.0$ sccm was characterized using cross-section transmission electron microscopy (TEM, Tecnai G2 F20 S-Twin) operated at 220 kV. Standard lift-out and thinning techniques were used for sample preparation to achieve electron transparency. Resistivity measurements were conducted using a four-probe method with a Physical Property Measurement System (PPMS-9, Quantum Design, Inc.). The temperature-dependent resistivity was measured during both cooling and heating processes. The phase transition temperature was also determined by analyzing resistivity data. Specifically, it was calculated as the average of the temperatures corresponding to the maximum changes in resistivity curves during cooling and heating cycles. This approach ensures an accurate determination of the phase transition temperatures based on detailed resistivity variations.

To investigate the impact of nitrogen vacancies on the structural phase transition and band structure of CrN(111) epitaxial thin films, first-principles calculations were performed using the Vienna Ab Initio Simulation Package (VASP). The calculations employed the generalized gradient approximation (GGA) with Perdew-Burke-Ernzerhof (PBE) functionals to account for exchange-correlation effects. The cubic unit cell of CrN was projected into a triangular lattice (R3m) to align with the (111) orientation of the films. In-plane lattice constants were fixed at the bulk value of $a$ = 2.93 Å, while out-of-plane lattice parameters were derived from temperature-dependent XRD measurements at 140 K and 300 K. Nitrogen vacancy concentrations were simulated at 0% and 8.33% to elucidate their influences. Initial structural optimizations were performed on a 2×2×1 supercell to identify the lowest energy configuration. This optimized structure was further expanded into a larger 2×2×1 supercell to accommodate the periodic requirement for the alternating double-layer antiferromagnets. Static calculations incorporated a Hubbard U correction of 3 eV. The impact of nitrogen vacancies on the structural phase transition is assessed by calculating the energy difference between the unit cell configurations after the transition (at 150 K) and before the transition (at 300 K), using this energy difference as a metric for evaluation. A plane-wave cutoff energy of 500 eV and an 8×8×7 Γ-centered k-point mesh ensured adequate Brillouin zone sampling, with convergence criteria set at $10^{-6}$ eV for total energy and 0.001 eV/Å for atomic forces.

## 3. Results and discussions

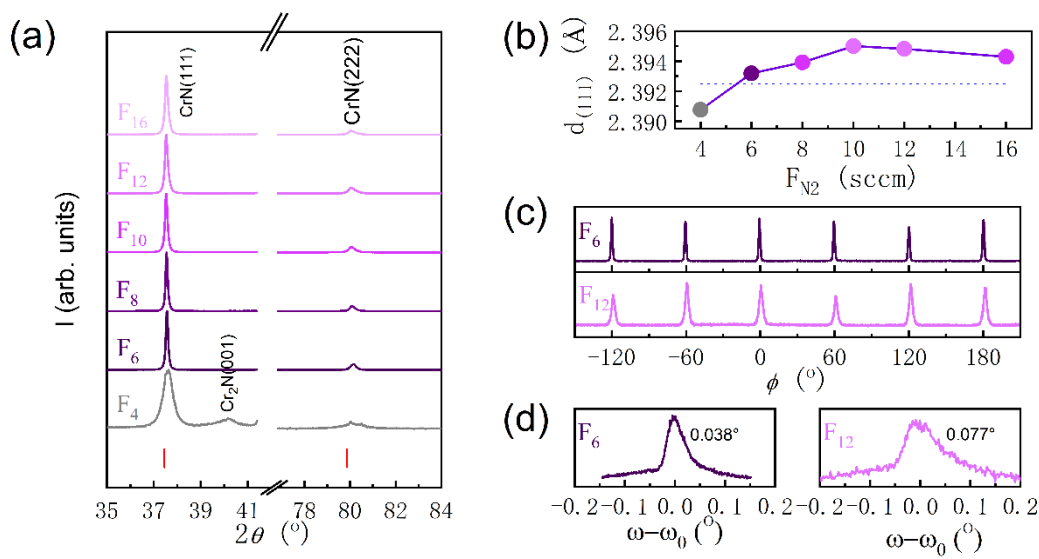

**Fig.2** Structural characterizations of CrN(111) epitaxial films at room temperature. (a) XRD spectra of CrN films prepared under different $F_{N2}$. (b) Out-of-plane interplanar spacing determined from the CrN(111) peaks, with the dotted line indicating the interplanar spacing of the CrN(111) planes in bulk powder. (c) φ-scans and (d) ω-rocking curve scans for films $F_{6.0}$ and $F_{12.0}$ over the (111) reflection.

The XRD patterns of films grown under different $F_{N2}$ are depicted in Fig. 2(a). For sample $F_{4.0}$, diffraction peaks at 34.47° and 72.86° correspond to the (111) and (222) planes of cubic CrN [4, 39], respectively, with a weak, broad peak at 40.16° indicating traces of $Cr_2N$(002). From $F_{6.0}$ to $F_{16.0}$, only the CrN (111) and (222) peaks are observed, suggesting highly oriented growth on $Al_2O_3$(0001) substrates. In Fig. 2(b), the out-of-plane spacing derived from the CrN(111) peaks increases from 2.3908 Å for $F_{4.0}$ to a maximum of 2.3950 Å at $F_{10.0}$, then gradually decreases to 2.3943 Å at $F_{16.0}$. This trend likely reflects the balance between chemical stress due to N vacancy variations and epitaxial strain from the substrate. Figure 2(c) shows φ-scan results for $F_{6.0}$ and $F_{12.0}$ over CrN(111), presenting sharp peaks separated by 60°, confirming epitaxial growth and indicating biaxial epitaxy associated with stacking faults [37, 47]. The ω-rocking curve scans in Fig. 2(d) reveal full widths at half maximum (FWHM) of 0.04° and 0.08° for $F_{6.0}$ and $F_{12.0}$, respectively, highlighting high crystalline quality. The narrower FWHM for $F_{6.0}$ may be attributed to lower $N_2$ gas flow rates, which increase N vacancies, relieve compressive strain, and enhance film quality. These findings indicate that CrN(111) epitaxial films can be successfully grown over a wide range of $N_2$ gas partial pressures.

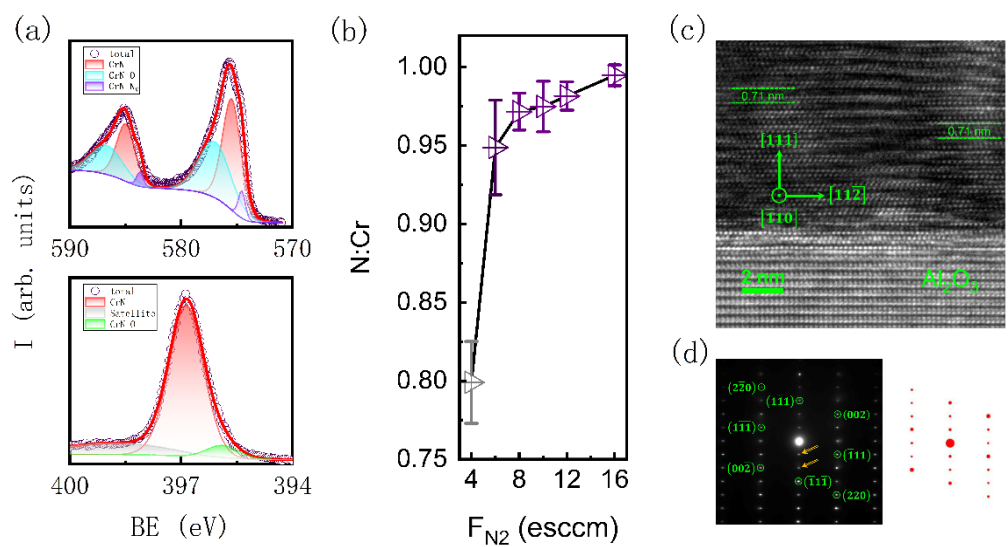

**Fig. 3** Composition and microstructure of CrN(111) epitaxial films. (a) Core-level spectra of Cr 2p and N 1s for sample $F_{6.0}$. (b) Atomic ratio of N to Cr, calculated from the integrated areas under the N 1s and Cr 2p core-level spectra. (c) Cross-sectional HR-TEM image taken along the $[\bar{1}10]$ zone axis for $F_{6.0}$. The observed distortion in lattice fringes and the appearance of "stripe" features suggest the presence of ordered nitrogen vacancies, with the period of these "stripes" being three times that of the CrN(111) plane spacing. (d) Selected area electron diffraction pattern of $F_{6.0}$. Green circles on the left highlight the diffraction spots expected for an ideal CrN structure,

while two additional diffraction spots, indicated by orange arrows, appear along the films normal direction. For comparison, a simulated diffraction pattern based on a nitrogen-deficient CrN supercell is shown on the right in red.

XPS measurements were carried out on the films to investigate the chemical state and composition of the films. As an example, the core-level spectra of Cr 2p and N 1s for sample $F_{6.0}$ are shown in Fig. 3(a). The binding energies for Cr 2p3/2 and N 1s are approximately 575.95 eV and 396.82 eV, respectively, consistent with previous reports for CrN [32, 48, 49]. The variation in nitrogen content with $F_{N2}$ was semi-quantitatively determined by analyzing the peak areas of N 1s and Cr 2p$_{3/2}$. The results in Fig. 3(b) indicate that the atomic ratio of N to Cr increases from 0.80 for $F_{4.0}$ to 0.95 for $F_{6.0}$, and reaches approximately 1.0 for $F_{16.0}$, corroborating the trend of increasing nitrogen content with higher $N_2$ gas flow rates. Notably, $F_{6.0}$ has the highest concentration of nitrogen vacancies, approximately 5.0 at%.

Cross-sectional HR-TEM characterization was performed on sample $F_{6.0}$ along the [$\bar{1}$10] zone axis to investigate the impact of nitrogen vacancies on the films microstructure. As shown in Fig. 3(c), a sharp interface between the film and substrate indicates epitaxial growth from the initial stage. The lattice fringes exhibit significant disorder, intertwining to form numerous "stripes" parallel to the film's surface. These stripes are spaced approximately 0.71 nm apart, roughly three times the interplanar spacing of CrN(111). According to the literature, these "stripes" signify stacking faults caused by ordered nitrogen vacancies [37, 44], and are responsible for the 60° intervals observed in the φ-scans (Fig. 2(c)). The selected area electron diffraction pattern on the left of Fig. 3(d) highlights the diffraction spots of an ideal CrN unit cell, marked by green circles. Along the normal direction of the film, two additional diffraction spots appear between each pair of CrN diffraction spots, indicated by orange arrows. These extra spots correspond to the "stripe" structures characterized by a three-fold lattice period. Furthermore, simulations using a distorted CrN unit cell that incorporates nitrogen vacancies accurately reproduce the observed diffraction patterns. Therefore, these TEM results corroborate the XPS findings, confirming the presence of substantial nitrogen vacancies in CrN films grown under low $N_2$ gas flow rates.

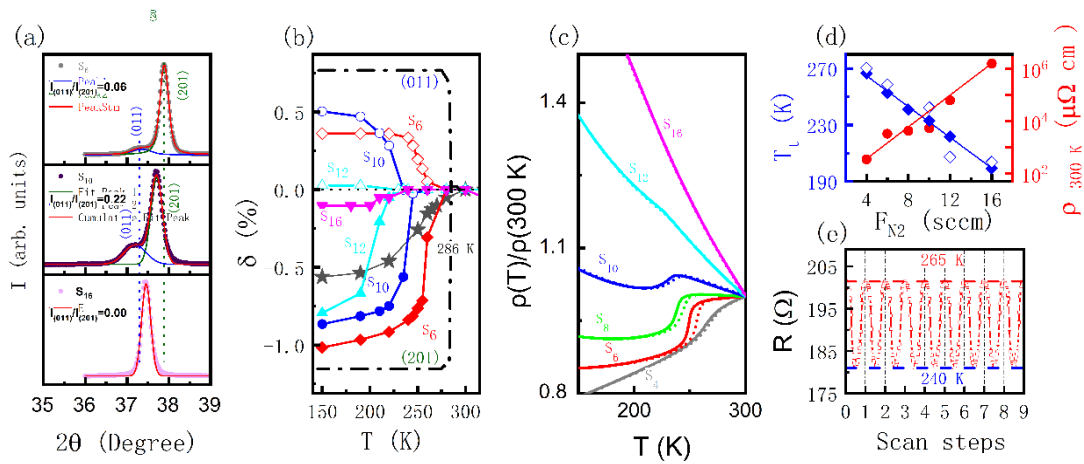

**Fig. 4** Characterizations of the phase transition of CrN(111) epitaxial films. (a) XRD patterns of samples $F_{6.0}$, $F_{10.0}$, and $F_{16.0}$ samples measured around the cubic CrN(111) plane at 150 K. Orthorhombic (011) and (201) peaks are fitted with blue and green lines, with their intensity ratio is provided in the inset. (b) Variation of the normalized changes of the out-of-plane interplanar spacing, $\delta$, with temperature. Hollow and solid symbols correspond to data for orthorhombic (011) and (201) planes, respectively. For comparison, results for CrN bulk powder are shown as black dash-dot lines [3]. (c) Normalized resistivity curves, $\rho(T)/\rho(300\ K)$, of CrN films with solid and dashed lines

representing cooling and heating processes. (d) Phase transition temperature ($T_N$, left axis) and room-temperature resistivity (right axis) vs. $F_{N2}$. The phase transition temperatures are derived from temperature-dependent XRD measurements (blue hollow symbols) and resistivity curves (blue solid symbols). (e) Resistance cycling test of $F_{6.0}$ between 265 and 240 K during cooling and heating.

The XRD spectra around the cubic CrN(111) plane for samples $F_{6.0}$, $F_{10.0}$, and $F_{16.0}$ at 150 K are presented in Fig. 4(a) (additional details in Fig. S2). The emergence of orthorhombic CrN (011) and (201) peaks in $F_{6.0}$ and $F_{10.0}$ indicates a first-order phase transition, while the single peak observed in $F_{16.0}$ suggests suppression of this structural transition. In $F_{6.0}$, both orthorhombic (011) and (201) peaks are positioned at higher angles compared to those of $F_{10.0}$, indicating significantly smaller out-of-plane interplanar spacings. Additionally, the intensity ratio of these two peaks is 0.06 for $F_{6.0}$, markedly lower than the 0.22 measured for $F_{10.0}$. These observations provide direct evidence supporting the hypothesis illustrated in Fig. 1(f), suggesting that N vacancies facilitate the contraction of triangular Cr atomic layers along the normal direction. Figure 4(b) shows the normalized changes of the out-of-plane interplanar spacing ($\delta = \frac{d(t) - d(300\ K)}{d(300\ K)} \times 100\%$) versus temperature for CrN(111) epitaxial films prepared under varying $F_{N2}$. Unlike most single-phase CrN(111) films, $F_{4.0}$ exhibits a less pronounced reduction in out-of-plane spacing around 280 K due to the presence of $Cr_2N$. For $F_{6.0}$, the cubic CrN(111) plane sharply splits into orthorhombic (011) and (201) planes near 260 K, with the contraction of triangular Cr atomic layers (0.8%) reaching up to 60% of bulk powders (1.2%), an order of magnitude greater than literature reports [21]. As $F_{N2}$ increases, the phase transition shifts to lower temperatures, and the reduction in out-of-plane spacing diminishes. By $F_{16.0}$, the splitting becomes indistinguishable, demonstrating that higher nitrogen deficiency promotes a more distinct structural phase transition in CrN(111) epitaxial films.

The evolution of the normalized resistivity curves, $\rho(T)/\rho(300\ K)$, with varying $F_{N2}$, as shown in Fig. 4(c), closely mirrors the results from variable temperature XRD measurements. Unlike the gradual decrease in resistivity near 270 K observed for mixed CrN and $Cr_2N$ phases of $F_{4.0}$ [38], film $F_{6.0}$ exhibits a significant drop in resistivity and observable thermal hysteresis around 260 K, indicating a dramatic structural and electronic phase transition [2, 50]. As $F_{N2}$ increases, the phase transition-induced resistivity jump and thermal hysteresis diminish, becoming negligible at $F_{N2}$ = 16.0 sccm. Concurrently, the structural phase transition temperature decreases steadily. Specifically, as illustrated on the left axis of Fig. 4(d), the structural phase transition temperatures decline linearly from 273 K to 193 K, crossing a span of 80 K, as $F_{N2}$ increases from 4.0 sccm to 16.0 sccm. Contrary to intuition, these results indicate that nitrogen vacancies contribute to an increasing the phase transition temperature of CrN(111) epitaxial films. Providing a reasonable explanation for this phenomenon is necessary, especially considering that the phase transition in CrN is driven by magnetic stress [8, 51]. Clearly, variations in phase transition temperatures are intimately linked to changes in internal stress due to nitrogen vacancies. As presented in Fig. 1(f), these vacancies reduce the inherent tensile stress between triangular Cr atomic layers. In nitrogen deficient CrN(111) films, the structural phase transition from low to high temperature primarily involves the expansion of these Cr layers. Consequently, lattice vibrations must compensate for the reduced tensile stress before initiating the phase transition, necessitating enhanced anharmonic effects and thus having higher phase transition temperature. Therefore, the phase transition temperature decreases with increasing $F_{N2}$. Considering the off-stoichiometric nature of CrN, reported variations in phase transition temperatures across studies likely correlated with differing N vacancy concentrations.

Moreover, the films become increasingly insulating with increasing $F_{N2}$. As shown on the right axis of Fig. 4(d), the resistivity at 300 K increases exponentially from $3.44\times10^2$ to $1.56\times10^6$ μΩ·cm, spanning more than three orders of magnitude. Crossing $F_{N2}$ from low to high temperatures, the resistivity curves in Fig. 4(c) evolve from a metal-to-metal transition at $F_{6.0}$, through a metal-to-insulator transition around $F_{8.0}$, to an insulator-to-insulator transition at $F_{10.0}$, with this transition ultimately disappearing by $F_{16.0}$. This sequence encompasses all documented types of electronic phase transitions and underscores that the nitrogen vacancy concentration dominates the electronic phase transition behavior of the films as well[16, 26, 30, 37]. To assess the reliability of the phase transition in nitrogen deficient CrN(111) epitaxial films, resistance cycling measurements were conducted for film of $F_{6.0}$ between 240 and 265 K. As shown in Fig. 4(e), no changes were observed after nine cycles, indicating that N vacancies do not compromise the stability of the structural phase transition. This robustness ensures the applicability of N-deficient CrN(111) epitaxial films in electronic devices.

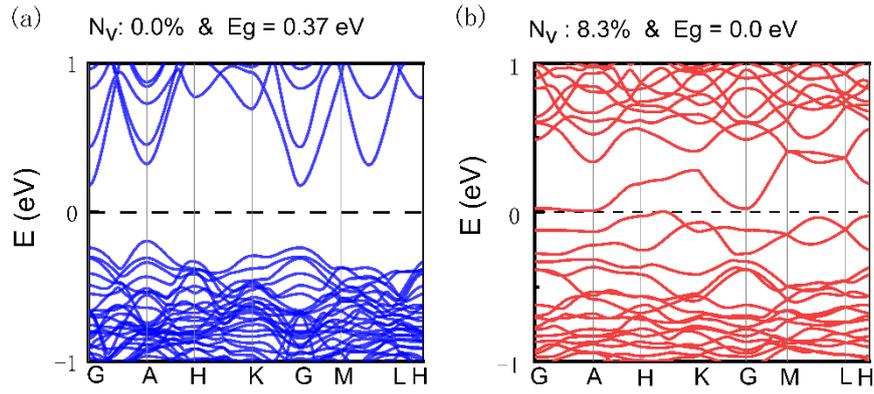

**Fig. 5** Influence of nitrogen vacancies on the band structure of stoichiometric and nitrogen deficient CrN(111) epitaxial films. The band structures for (a) stoichiometric CrN (space group: R3m), and (b) off-stoichiometric CrN with 8.3% nitrogen vacancy concentration.

Simulations were performed using VASP to further investigate the impact of nitrogen vacancies on the structural phase transition and band structure of CrN(111) epitaxial thin films. The effect on structural phase transition was evaluated by calculating the energy difference between the superlattice after (150 K) and before (300 K) the phase transition. For stoichiometric CrN, this energy difference, $\Delta E = E(150\ K) - E(300\ K)$, was 0.005 eV/f.u., whereas, for a nitrogen vacancy concentration of 8.3%, it decreased to -0.23 eV/f.u., indicating significant triangular Cr atomic layer contraction due to nitrogen vacancies. These findings align with literature data [44], and observed trends in the Cr-N system, where the lattice constant of triangular Cr layers increases from 2.21 Å in $Cr_2N$ to 2.39 Å in CrN with increasing nitrogen content. As shown in Fig. 5(a), our calculations show that stoichiometric CrN exhibits a bandgap of 0.37 eV, characteristic of semiconducting behavior, consistent with theoretical and experimental results [8]. In contrast, at 8.33% nitrogen vacancy concentration of Fig. 5(b), the bandgap closes as the nitrogen vacancies induce a transformation from the polar covalent bonds of "Cr-N-Cr" to metallic bonds. The observed bandgap closure due to nitrogen vacancies explains the trend in resistivity shown in Fig. 4(c), where resistivity decreases with decreasing nitrogen vacancy concentration. Specifically, as nitrogen vacancy concentration decreases, the conductivity of CrN(111) films diminishes. Overall, these theoretical calculations not only support the experimental hypothesis proposed in Fig. 1 but also effectively explain subsequent experimental measurements, highlighting the significant influence of nitrogen vacancies on the phase transition of CrN(111) epitaxial films.

## 4. Conclusion

This study identifies that the internal stress resulting from the splitting of cubic CrN {111} triangular planes into orthorhombic {011} and {201} planes with differing spacings inhibits structural phase transitions. We propose that this internal stress can be alleviated by the chemical stress induced by nitrogen vacancies, thereby facilitating phase transitions in CrN(111) epitaxial films. To explore this idea, we grew CrN(111) epitaxial films with varying nitrogen vacancy concentrations on $Al_2O_3$(0001) substrates by adjusting nitrogen flow rates. Experimental results show that nitrogen vacancies significantly enhance the vertical contraction of the films, reaching up to 60% of CrN powders, and increasing the phase transition temperature by 80 K. Additionally, decreasing nitrogen vacancy concentration increases film resistivity by approximately four orders of magnitude, shifting electronic transitions from metal-to-metal to insulator-to-insulator. First-principles calculations confirm that nitrogen vacancies do facilitate the contraction of cubic CrN(111) triangular planes and cause bandgap closure, leading to metallic behavior. In summary, our findings indicate that nitrogen vacancy concentration governs the phase transition dynamics of CrN films, enhances understanding of CrNs phase transition mechanisms, and highlights the potential for turning these transitions via nitrogen vacancy modulation, thus paving the way for developing high-performance CrN-based electronic devices.

# Acknowledgements

This work was supported by the "333 Talent Project" of Hebei province (grant No. C20231105) and the Science Foundation of Hebei Normal University, China (grant No. L2024B08), and the National Natural Science Foundation of China (grant no. 51901067, 51971087, 52101233, and 52071279), the Natural Science Foundation of Hebei Province (grant no. E2019205234), the Science and Technology Research Project of Hebei Higher Education (grant no. QN2019154), and the Science Foundation of Hebei Normal University (grant no. L2019B11).

**The supplementary materials**

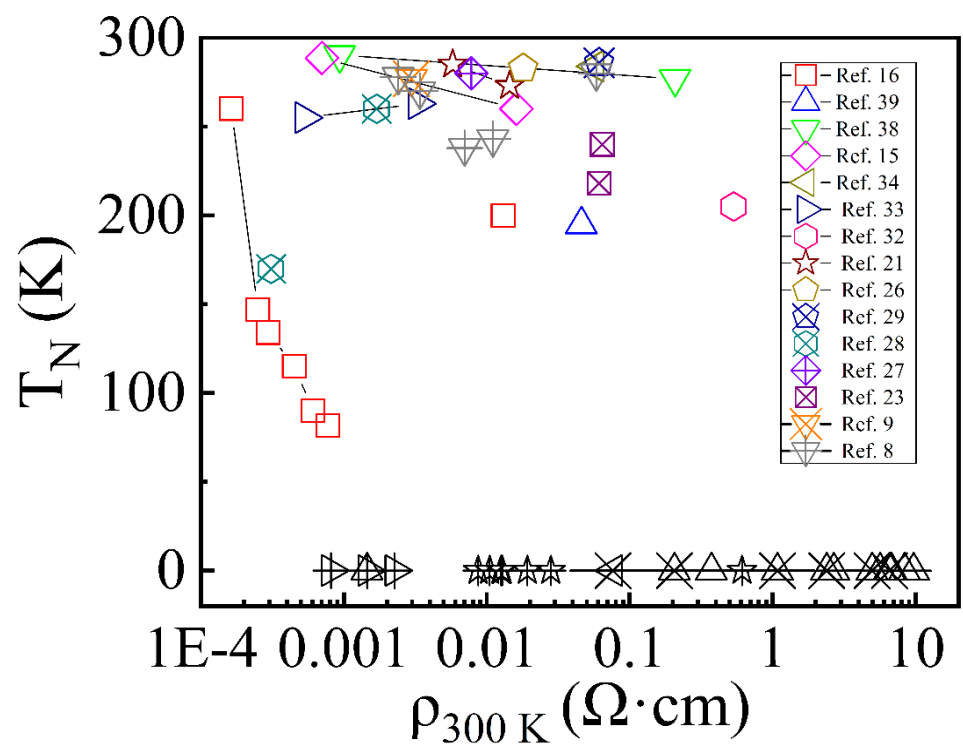

**Fig. S1.** Relationship between the reported phase transition temperatures and room-temperature resistivity in CrN thin films. For samples exhibiting no phase transitions, the phase transition temperature is designated as zero.

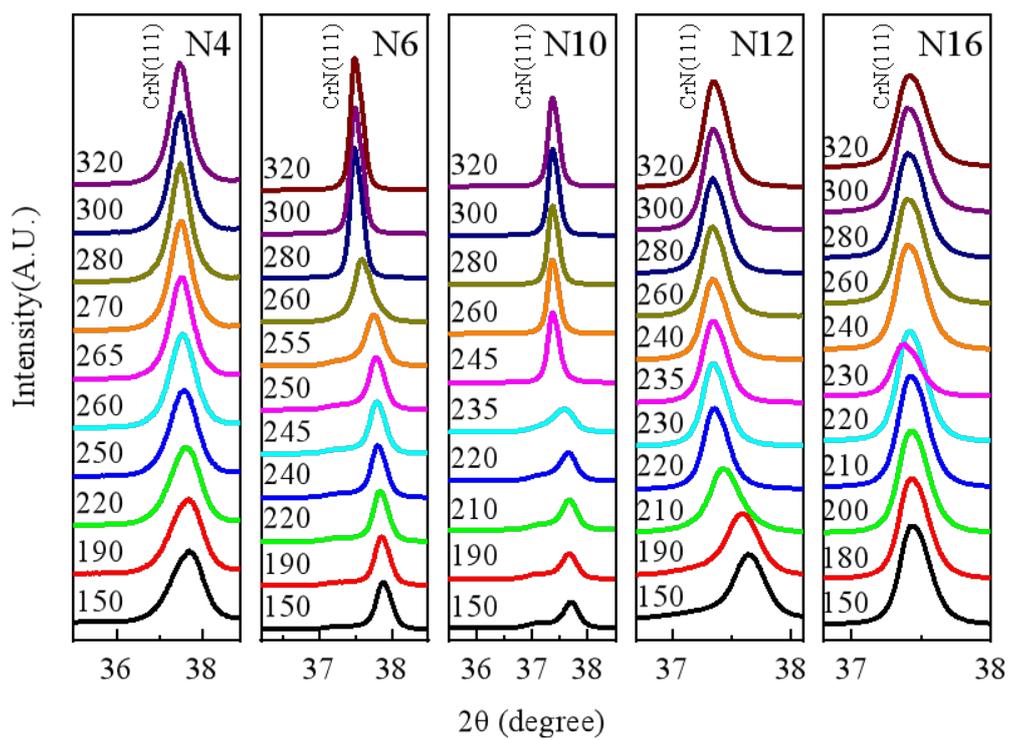

**Fig. S2.** Variable temperature XRD patterns of CrN(111) films at nitrogen flow rates of 4.0, 6.0, 10.0, 12.0, and 16.0 sccm.